\documentclass[aps,superscriptaddress]{revtex4}

\usepackage{graphicx}
\usepackage{bm}
\usepackage{amssymb}
\usepackage{amsmath}
\usepackage{dcolumn}
\usepackage[latin1]{inputenc}
\usepackage[english]{babel}
\usepackage{physics} 

\begin{document}
\renewcommand{\arraystretch}{1.2}

\title{A quark core-gluon model for heavy hybrid baryons}

\author{Lorenzo \surname{Cimino}}
\email[E-mail: ]{lorenzo.cimino@umons.ac.be}
\thanks{ORCiD: 0000-0002-6286-0722}
\affiliation{Service de Physique Nucl\'{e}aire et Subnucl\'{e}aire,
Universit\'{e} de Mons,
UMONS Research Institute for Complex Systems,
Place du Parc 20, 7000 Mons, Belgium}

\author{Cintia T. \surname{Willemyns}}
\email[E-mail: ]{Cintia.Willemyns@vub.be}
\thanks{ORCiD: 0000-0001-8114-0061}
\affiliation{Service de Physique Nucl\'{e}aire et Subnucl\'{e}aire,
Universit\'{e} de Mons,
UMONS Research Institute for Complex Systems,
Place du Parc 20, 7000 Mons, Belgium}
\affiliation{Vrije Universiteit Brussel, Pleinlaan 2, 1050 Brussels, Belgium}

\author{Claude \surname{Semay}}
\email[E-mail: ]{claude.semay@umons.ac.be}
\thanks{ORCiD: 0000-0001-6841-9850}
\affiliation{Service de Physique Nucl\'{e}aire et Subnucl\'{e}aire,
Universit\'{e} de Mons,
UMONS Research Institute for Complex Systems,
Place du Parc 20, 7000 Mons, Belgium}
\date{\today}

\begin{abstract}
\textbf{Abstract} Besides the ordinary hadrons, QCD allows the existence of states in which excitations of the gluonic field can play the role of valence particles, either alone in a glueball, or coupled to quarks in a hybrid. So, hybrid baryons, made of three quarks and a gluon, can a priori exist. Till now, there is no experimental evidence for such exotic hadrons but experimental efforts are being made to search for them at CEBAF Large Acceptance Spectrometer. In this work, a hybrid baryon is considered as a two-body system composed of a color octet three-quark core and a gluon, interacting via a QCD-inspired interaction. A semirelativistic potential model is built in which the dominant interaction is a potential simulating the flux tube confinement, and the Casimir scaling is assumed to link interactions between triplet and octet color sources. This picture is similar to the quark-diquark description for baryons. It is chosen in order to take properly into account the helicity of the gluon. Only $cccg$ and $bbbg$ states are considered because the strong mass asymmetry between the quark core and the gluon is expected to favor the formation of the core. As the results for heavy hybrid baryons seem relevant, we consider this paper as a proof of concept which can be extended for the study of light hybrid baryons.
\end{abstract}

\maketitle

\section{introduction}
\label{sec:intro}

Nowadays, good experimental candidates exist for exotic hadrons \cite{pdg22,chen23}. Besides mesons, baryons and multiquarks states, QCD allows also the existence of states in which excitations of the gluonic field play the role of valence particles, either alone in a glueball, or coupled to quarks in a hybrid. So, hybrid baryons, made of three quarks and a gluon, can a priori exist. These exotic hadrons have already been studied with various models: bag model \cite{barn83}, large-$N_c$ approach \cite{chow99}, flux-tube model \cite{page05}, lattice QCD \cite{dude12}, and QCD sum rules \cite{aziz18}. Unfortunately, though these models predict the existence of hybrid baryons, their predictions for masses and structures differ considerably from each other. 

From the experimental side, there is no clear signal yet as to the existence of these hybrid baryons. But, at present, experimental efforts are being made to search for hybrid baryons at CEBAF Large Acceptance Spectrometer (CLAS) in the experimental Hall B at Jefferson Lab \cite{pac44,burk18}. The experiment has been delayed, but the first data will be available in the next few years. Identifying hybrid baryons will be more difficult than hybrid mesons, as the latter can have exotic quantum numbers that are forbidden for states containing only constituent quarks. However, this is not the case for hybrid baryons which have quantum numbers that are also populated by ordinary baryons. So, mixings are possible between hybrid baryons and excited three-quark states. Hybrid baryons should then appear in terms of overpopulation with respect to some models of baryon excitations. Fortunately, the nature of the states produced at CLAS can be explored by investigating the $Q^2$ dependence of the resonance coupling in electroproduction processes \cite{pac44,burk18}. Differences with ordinary baryons are expected due to the additional gluonic component in the wave function of hybrids. For the same reason, decay products of hybrid baryons must differ from decay products of ordinary baryons.

The differences between the various models of hybrid baryons and their possible detection in the near future make that a better understanding of these objects is of the utmost importance for the correct identification of these new states. This paper is intended to be a first step into the development of a reliable model of hybrid baryons based on a semirelativistic potential approach (with a relativistic kinetic energy but without full covariance) \cite{close,luch91}. Potential models have been successfully used to study masses and static properties of ordinary hadrons \cite{godf85,caps86} as well as more exotic states \cite{math08,carl08,gian19,buis07}. So, it seems relevant to challenge this approach for hybrid baryons. Results obtained can shed new light on these exotic states. If lattice QCD computations are probably the best way to investigate the spectra of hadrons, potential models allow to explore in detail the various aspects of the strong interaction.

Even if heavy hybrid baryons will not be the first exotic hadrons to be produced in planned experiments \cite{pac44,burk18}, they will be the subject of this work because they are simpler to study, as it will be explained below. In our model, a hybrid baryon is assumed to be a two-body system composed of an inert color octet core of three quarks and a constituent gluon, both interacting via a QCD-inspired potential. The reasons for the choice of this model are described in Sec.~\ref{sec:constHeli}. For short, these are the helicity quantum number of the gluon, because of its vanishing mass, and the big mass asymmetry between the two effective constituents. Despite the simplicity of the model, we hope to catch the main physical properties of heavy hybrid baryons. Masses, sizes and quantum numbers of the color octet quark core are computed in Sec.~\ref{sec:qcore}. Sec.~\ref{sec:ggandcg} is devoted to the description of the quark core-gluon interaction. Masses and quantum numbers of the lowest $cccg$ and $bbbg$ hybrid baryons are computed in Sec.~\ref{sec:MHHB}. Some concluding remarks and possible outlook are presented in Sec.~\ref{sec:conclu}. Two appendices give respectively general formulas about two-body helicity states and the Lagrange-mesh method to compute eigensolutions of two-body semirelativistic Hamiltonians.

\section{Potential models with helicity}
\label{sec:constHeli}

The gluon is a vector particle. Due to its massless character, it is characterized by a helicity, that is to say only two projections $\pm 1$ of its intrinsic angular momentum, and not by a spin with a third projection $0$. In \cite{math08}, it is shown that a semirelativistic potential model of two-gluon glueballs is possible provided that the helicity of gluons is correctly taken into account. This is confirmed in \cite{buis09}. So, it is expected that the same constraint must be imposed on a semirelativistic potential model of hybrid baryons. If the helicity formalism is well known and manageable in practice for two-body systems \cite{jaco59}, this is not true for three particles \cite{wick62,berm65} or more  \cite{werl63}, for which a lot of work must be done to make the theory usable in potential models. In order to keep a correct treatment of the helicity for a hybrid baryon manageable, the particle is considered here as a two-body system formed by two color octet sources: a point-like massless gluon with helicity interacting with an inert extended core of quarks in its ground state with a defined spin. The quark core is described as inert because no excitation is allowed. This simplification will be discussed in Sec.~\ref{sec:conclu}. This model is similar to the quark-diquark description of baryon in which two quarks form a cluster interacting with the third one. This model has a long history but it is still quite popular nowadays \cite{flec88,gian09}. This description is interesting to study the internal structure of baryons, but it is also used to compute properties of multiquark systems as tetraquarks \cite{carl08} and pentaquarks \cite{gian19}. The quark-diquark structure in a baryon is favored by the presence of two heavy quarks forming a very tied cluster in its ground state which interacts with a third quark lighter than the other ones \cite{flec88,gian09}. That is why we focus in this paper on hybrid baryons containing three heavy quarks in order to maximize the probability of formation of a strongly tied quark core, very little disturbed by the dynamics of the massless gluon. 

\section{Quark core description}
\label{sec:qcore}

As excitations are generally better reproduced than absolute masses, we will focus on the mass difference between heavy hybrid baryons and the ground state of the ordinary baryon made with the same quark content as the one of the quark core. In this paper, we only consider clusters with three identical heavy quarks, $qqq = ccc$ or $bbb$. The semirelativistic three-body problem can be accurately solved by an expansion in oscillator bases \cite{nunb77,silv00,chev23}. All quantities are given in natural units $\hbar = c = 1$.

\subsection{Ordinary baryons}
\label{sec:OB}

With three identical  quarks, the flavor state of a baryon is completely symmetrical. As we consider only the system in its ground state with zero total orbital angular momentum $L_B$, the spatial wave function is also completely symmetrical. For a baryon, that is to say a colorless state completely antisymmetrical, this implies a completely symmetrical spin state $S_B=3/2$ in order that the total wave function be completely antisymmetrical. So, baryons used as references are characterized by $J_B^{P_B}=3/2^+$. Our Hamiltonian relies on the simple Cornell potential for mesons developed in \cite{fulc94}. No spin effect is taken into account within this model, but such a contribution is  small with respect to those coming from the spin-independent central part of the interaction. With the notations of \cite{fulc94}, the $q\bar q$ potential for a meson is written
\begin{equation}
\label{Vqqbar}
V_{q\bar q} = A\,r -\frac{\kappa}{r}.
\end{equation}
A precise meaning must be given to parameters $A$ and $\kappa$ in order to extend this interaction for three-quark systems. The linear term is a good approximation for the energy of the flux tube in a meson and $A$ is the string tension (or the linear energy density). It is natural to interpret the Coulomb term as due to the one-gluon exchange. In this case, $\kappa$ is assumed to be given by $\langle \bm F_q \cdot \bm F_{\bar q} \rangle\, \alpha_S = -(4/3)\,\alpha_S$, where $\alpha_S$ is the strong coupling constant and $\bm F_i \cdot \bm F_j$ is the usual two-body color operator. Then $\alpha_S =(3/4) \kappa = 0.328$. For heavy quarks, this potential has also the advantage not to necessitate the presence of constants whose possible color dependence is not clear and thus not easily transposable to three-body systems. A nonrelativistic and a semirelativistic versions exist and both provide a good description of the spin centers of gravity of a large set of heavy and light mesons. The values of the parameters for the semirelativistic version used in this work are given in Table~\ref{tab:paramq}. It is worth noting that there is only a gap of about 10-20~MeV between the masses of three-quark systems computed with the nonrelativistic and the semirelativistic versions. But we prefer to use the semirelativistic one to be coherent with the Hamiltonians associated with systems containing gluons.

\begin{table}[htb]
\begin{center}
    \begin{tabular}{clccl}
        \hline\hline 
$m_c$ & 1.320 GeV & \ & $A$ & 0.203 GeV$^2$  \\
$m_b$ & 4.731 GeV & \ & $\kappa$ & 0.437 \\
 & & \  & $f$ & 1.086 \\
        \hline\hline
    \end{tabular} 
    \caption{Parameters for baryon and quark-core Hamiltonians~(\ref{HB}) and (\ref{HC}) \cite{fulc94,silv04}. \label{tab:paramq}}
\end{center}
\end{table}

The Hamiltonian for a baryon with three identical quarks with a mass $m_q$ is then given by
\begin{equation}
\label{HB}
H_B=\sum_{i=1}^3 \sqrt{\bm p_i^2+m_q^2} + \frac{1}{2} \sum_{i<j}^3 \left( f A\,r_{ij} -\frac{\kappa}{r_{ij}} \right),
\end{equation}
with the usual notation $r_{ij} = |\bm r_i - \bm r_j|$. The $1/2$ factor in the potential part has two different origins. For the Coulomb part, $\langle \bm F_q \cdot \bm F_q \rangle=\langle \bm F_q \cdot \bm F_{\bar q} \rangle/2=-2/3$ in a baryon. The more realistic confinement in a baryon is the Y-junction, that is to say three flux tubes, each generated by quark, connecting into a point $\bm r_0$ minimizing the potential energy. As this interaction is very difficult to treat, it is replaced here by a good two-body approximation 
\begin{equation}
\label{Vconf}
A \min_{\bm r_0}\sum_{i=1}^3 |\bm r_i - \bm r_0| \approx A\frac{f}{2} \sum_{i<j}^3 r_{ij},
\end{equation}
with $f=1.086$ \cite{silv04}. The ground state masses of $ccc$ and $bbb$ are given in Table~\ref{tab:mBC}. No experimental data are available for a comparison. 

\subsection{Octet quark core}
\label{sec:OC}

For the $L_C=0$ color octet $qqq$ quark-core in its ground state, the flavor-space wave function is identical to the one for the baryon and so completely symmetrical. As the color octet state is a mixed symmetric one, it must be combined with a spin state with a mixed symmetry, that is to say $S_C=1/2$, in such a way that the total wave function be completely antisymmetrical. If $\chi$ is a 3-body spin-$1/2$ state and $\phi$ a 3-body color octet state, the completely antisymmetrical spin-color state is written \cite{close}
\begin{equation}  
\label{spincol}
\frac{1}{\sqrt{2}} \left( \chi^\textrm{MS} \phi^\textrm{MA} - \chi^\textrm{MA} \phi^\textrm{MS} \right),
\end{equation}
where MS (MA) stands for Mixed Symmetric (Mixed Antisymmetric). So, quark cores under study are characterized by $J_C^{P_C}=1/2^+$. 

\begin{figure}[htb]
\includegraphics[width=12.68cm,height=7cm]{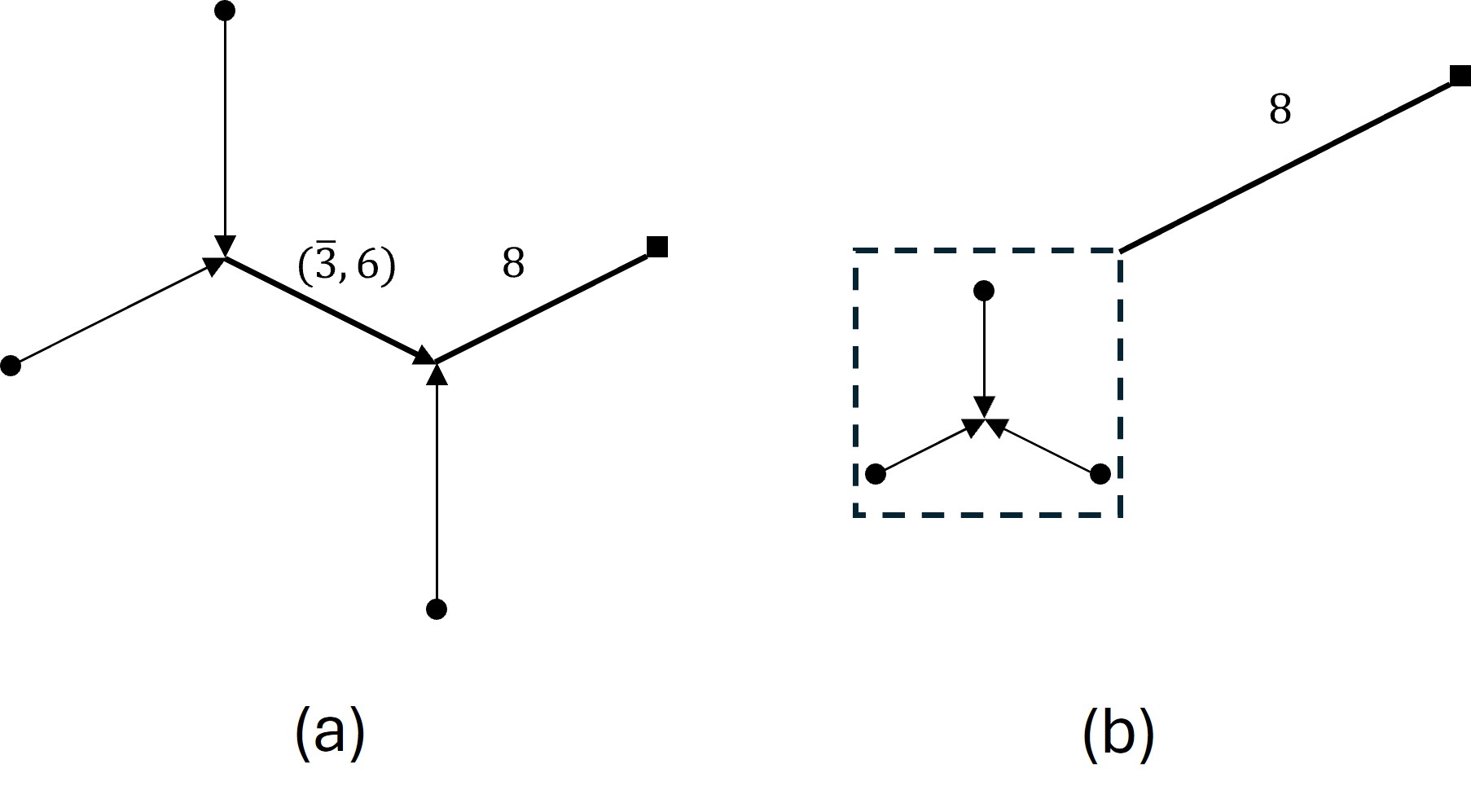}
\caption{Color flux-tube structures for a hybrid baryon with three quarks (circles) and a gluon (square): (a) full description; (b) quark core-gluon approximation. The color representations of non-fundamental flux-tubes are indicated. \label{fig:flux}}
\end{figure}

In the flux tube picture of the confinement, the dominant interaction for the three quarks and the gluon can be described by the part (a) of Fig.~\ref{fig:flux} \cite{deng13}. This is a very complicated interaction which is not relevant in the framework of our model. So, it is replaced by the configuration presented on the part (b). In this case, each quark produces a flux tube with the same energy density than in a baryon, and the connection into a color octet configuration is possible thanks to the color neutralization achieved by the octet flux tube generated by the gluon (see Sec.~\ref{sec:ggandcg}). For the Coulomb part, it is necessary to compute the value of $\langle \bm F_q \cdot \bm F_q \rangle$ for the $qq$ pairs. The mean value is $1/3$ for the symmetric representation $\bm 6$ and $-2/3$ for and the antisymmetric representation $\bm \bar{\bm 3}$. Finally, the Hamiltonian for the quark core is written
\begin{equation}
\label{HC}
H_C=\sum_{i=1}^3 \sqrt{\bm p_i^2+m_q^2} + \frac{1}{2} \sum_{i<j}^3 \left( f A\,r_{ij} - \frac{1}{4} \frac{\kappa}{r_{ij}} \right),
\end{equation}
where the factor $1/4$ comes from the mean value of $\langle \bm F_q \cdot \bm F_q \rangle$ for the spin-color state~(\ref{spincol}).

The quark core is not a point-like particle. It is thus necessary to compute the density of color (that is to say the density of quarks) inside the core to compute the correct quark core-gluon interaction (see Sec.~\ref{sec:ggandcg}). A natural definition for a normalized $N$-body density is \cite{naza13} 
\begin{equation}
\label{densityN}
\rho(\bm r) = \frac{1}{N}  \sum_{j=1}^N \int\ldots\int \left| \psi \right|^2 \delta(\bm r-\bm r_j)\, d\bm r_1 \ldots d\bm r_N, 
\end{equation}
where $\psi$ is the $N$-body wave function. Accurate computations in oscillator bases show that the probability of the first component of the expansion (the product of two ground states of oscillator functions in Jacobi coordinates) amounts for more than 90\% of the wave function for ground state cores. Let us note that, with a trial state reduced to this unique component, the masses are reproduced with a relative error of 0.1\%. We decided to use this approximation to compute the density of the core that then takes the simple form 
\begin{equation}
\label{rho}
\rho(\bm r) = \frac{\lambda^3}{\pi^{3/2}} e^{-\lambda^2 r^2}. 
\end{equation}
This allows to keep an analytical form for the final quark core-gluon interaction (see Sec.~\ref{sec:ggandcg}). The parameter $\lambda$ is easily determined with the numerical method. Results for the masses of the ground states of $ccc$ and $bbb$ baryons and octet cores are given in Table~\ref{tab:mBC}. The reduction of the Coulomb attraction produces a significant increase for the quark core mass with respect to the corresponding baryon. This contributes to the mass of hybrid baryons. 

\begin{table}[htb]
\begin{center}
    \begin{tabular}{ccccc}
        \hline\hline
State & $m_B$ & $m_C$ & $\Delta$ & $\lambda$ \\
\hline
$ccc$ & 4.822 & 5.119 & 0.297 & 0.825 \\
$bbb$ & 14.401 & 14.894 & 0.493 & 1.261 \\
        \hline\hline
    \end{tabular} 
    \caption{Ground state masses for baryons ($m_B$) and color octet quark cores ($m_C$), mass gap $\Delta = m_C - m_B$, and size parameter $\lambda$ in (\ref{rho}) for quark cores (all quantities are in GeV). \label{tab:mBC}}
\end{center}
\end{table}

\section{Gluon-gluon and quark core-gluon interactions}
\label{sec:ggandcg}

Assuming that the Casimir scaling hypothesis is valid \cite{bali00,sema04}, the relation between the interaction $V_{gg}$ for two gluons and $V_{q\bar q}$, both for singlet color systems, is given by 
\begin{equation}
V_{gg}= \frac{9}{4} V_{q\bar q},
\end{equation}
where the factor $9/4$ is given by the ratio between the values of the quadratic QCD Casimir operator for octet ($\langle \bm F_g^2 \rangle=3$) and triplet ($\langle \bm F_q^2 \rangle=4/3$) configurations. That is the assumption used in \cite{math08} to build the semirelativistic Hamiltonian for a two-gluon glueball 
\begin{equation}
\label{Hgg}
H_{gg}=2 \sqrt{\bm p^2+m_g^2} + \frac{9}{4} \sigma \,r -3 \frac{\alpha_S}{r},
\end{equation}
where the notations of \cite{math08} are used: $m_g$ is the gluon mass and $\sigma$ is the string tension of the mesonic flux tube (parameter $A$ in potential~(\ref{Vqqbar})). $\langle \bm F_g \cdot \bm F_g \rangle=-3$ for a two-gluon glueball. Two sets of parameters are determined in \cite{math08}, a first one for gluon with a spin and a second one for gluon with a helicity. These parameters are gathered in Table~\ref{tab:paramg} and compared with the corresponding ones for potential~(\ref{Vqqbar}). The agreement between the glueball masses computed with (\ref{Hgg}) and the predictions from a lattice QCD calculation is only possible with two different values of $\alpha_S$, but the mass hierarchy is far better reproduced for the gluon with a helicity. One may wonder why test the spin status for a gluon when one knows very well that it must be false since the gluon is massless. This is actually because the gluon can gain an effective mass inside the hadron due to the confinement \cite{luch91}. So, it seems relevant to check if the gluon could lost its helicity to gain a spin. The conclusion of  \cite{math08} is that the helicity is preserved and must be used in a potential model. 

It could seem strange that different values for the string tension and the strong coupling constant are assigned according to the particles, quarks or gluons, considered. We preferred to keep different sets of parameters for two reasons. First, it is difficult to find common values which give good results for all systems, taking into account the simplicity of the models. Second, we have then the possibility to compare masses of hybrid baryons according to the choice of the spin or the helicity made for the gluon. 

\begin{table}[htb]
\begin{center}
    \begin{tabular}{clll}
        \hline\hline
 & spin & helicity & $q\bar q$\\
\hline
$m_g$ & 0 & 0  \\
$\sigma$ & 0.185 GeV$^2$ & 0.185 GeV$^2$ & 0.203 GeV$^2$ \\
$\alpha_S$ & 0.200 & 0.450 & 0.328 \\
        \hline\hline
    \end{tabular} 
    \caption{Parameters for glueball and hybrid baryon Hamiltonians \cite{math08}. Corresponding parameters for the $q\bar q$ systems taken from Table~\ref{tab:paramq} are also indicated  \label{tab:paramg}}
\end{center}
\end{table}

We assume the universality of the interaction between two color octet sources. So, the potential between the gluon and the color octet quark core is the same as the one between two gluons, since the internal structure of the core is irrelevant at this level, but with the difference that the quark core is not point-like. The interaction between two point-like sources must then be convoluted with the density of the extended source according to the formula \cite{gian09}
\begin{equation}
\label{Vconv}
\tilde V(\bm R) = \int \rho(\bm r)\, V\left( \left|\bm R + \bm r  \right| \right)\, d\bm r.
\end{equation}
Let us note that our definition of $\rho(\bm r)$ is different compared to the one in \cite{gian09}, but we must deal with three quarks, instead of two. No normalization is necessary in~(\ref{Vconv}) since $\rho(\bm r)$ is already normalized. Using relation~(\ref{rho}), the Hamiltonian for hybrid baryons is written
\begin{equation}
\label{HCg}
H_{H\!B}=\sqrt{\bm p^2+m_g^2} + \sqrt{\bm p^2+m_C^2} + \frac{9}{4} \sigma \left[ \frac{e^{-\lambda^2 r^2}}{\sqrt{\pi} \lambda}+ \left(r +\frac{1}{2\lambda^2 r}\right) \textrm{erf}(\lambda\,r) \right] -3\, \alpha_S\frac{\textrm{erf}(\lambda\,r)}{r},
\end{equation}
where $m_C$ is the quark core mass, $\lambda$ its size parameter computed with Hamiltonian~(\ref{HC}), and $\textrm{erf}$ is the usual error function. The eigensolutions for this Hamiltonian must be found for spin or helicity quark core-gluon states. With a purely central potential without orbital nor spin dependence, the main difficulty is to compute the action of the square of the relative angular momentum operator $\bm L^2$ contained in the operator $\bm p^2$. For helicity quark core-gluon states, this is a very similar problem to the one in \cite{math08}, but with the difference that couplings can appear between helicity states (see Sec.~\ref{sec:MHHBheli}).

\section{Masses of heavy hybrid baryons}
\label{sec:MHHB}

Accurate eigenvalues and eigenstates of Hamiltonian~(\ref{HCg}) can be easily computed with the Lagrange-mesh method \cite{sema01,baye15} (see Sec.~\ref{sec:LagMesh}). We consider two cases: a gluon with a spin and a gluon with a helicity. In both cases, masses $m_{H\!B}$ of states with the lowest $J^P$ numbers are computed. It seems to us more relevant to present the mass gap $m_{H\!B}-m_B$ with respect to the lowest corresponding baryon mass $m_B$ given in Table~\ref{tab:mBC}. We do not expect that results with spin are relevant for the physics, but we found interesting to compare them to the results with helicity, as it is done for two-gluon glueballs in \cite{math08}.

\subsection{Gluon with spin}
\label{sec:MHHBspin}

The parameters $\sigma$ and $\alpha_S$ are taken from the ``spin" column in Table~\ref{tab:paramg}. The octet quark core is a $J_C^{P_C}=1/2^+$ state while the gluon is a $J_g^{P_g}=1^-$ state. The total spin $S$ of a hybrid baryon is a good quantum number with $1/2$ and $3/2$ possible values. The relative angular momentum $L$ is also a good quantum number and the total parity is $P=(-1)^{L+1}$. A tower of degenerate $J^P=1/2^-$ and $3/2^-$ hybrid baryons can be formed for $L=0$. With $L=1$, another tower of degenerate $J^P=1/2^+$, $3/2^+$, and $5/2^+$ states is possible. The degeneracies appear because the interaction is purely central. Results for $cccg$ and $bbbg$ are given in Table~\ref{tab:mqqqgs}. The mass gap is nearly the same for both flavors.

\begin{table}[htb]
\begin{center}
    \begin{tabular}{cllll}
        \hline\hline
 $J^P$ & $L$ & $n_r$ & $cccg$ & $bbbg$ \\
\hline
  $1/2^-$, $3/2^-$ & 0 & 0 & 1.652 & 1.635 \\
  $1/2^+$, $3/2^+$, $5/2^+$ & 1 & 0 & 2.194 & 2.220 \\
  $1/2^-$, $3/2^-$ & 0 & 1 & 2.469 & 2.444 \\
        \hline\hline
    \end{tabular} 
    \caption{Mass gap $m_{H\!B}-m_B$ in GeV for some $cccg$ and $bbbg$ hybrid baryons for a gluon with a spin. The number $n_r = 0 \, (1)$ indicates the ground state (the first radial excitation) for the given value of $L$. \label{tab:mqqqgs}}
\end{center}
\end{table}

\subsection{Gluon with helicity}
\label{sec:MHHBheli}

With a helicity for the gluon, things are very different, not only because the parameters $\sigma$ and $\alpha_S$ are taken from the ``helicity" column in Table~\ref{tab:paramg}, but also because of the very different structure for the spin-orbital wave functions of hybrid baryons. If a spin could be assigned to the gluon, the two-body quark core-gluon states representing hybrid baryons should be ``ordinary'' two-body spin states usually noted $|^{2S+1}L_J\rangle$. As gluons are characterized by a helicity, the two-body states are in fact the states $|H_{\pm};J^P;\pm J_c\, 1\rangle$ built in Sec. \ref{sec:SpinHeli}, with a given spin $J_C$ for the quark core, in the same spirit of \cite{math08}. These helicity states can be expanded in states $|^{2S+1}L_J\rangle$ that are easier to manipulate
\begin{equation}
\label{HeliExpSpin}
|H_{\pm};J^P;\pm J_c\, 1\rangle = \sum_{L,S} C_{L,S,J}\,|^{2S+1}L_J\rangle .
\end{equation}
In general, $S$ and $L$ are not good quantum numbers. This decomposition is a pure mathematical trick to allow a direct computation of the action of various operators on the helicity states. We list here the states with the lowest $J^P$ quantum numbers that can be formed with a spin $J_C=1/2$. We use a simplified notation $|J^P,\alpha\rangle$ where the index $\alpha$ allows to distinguish the different possible internal structures. The value of $w_{\alpha\beta} = \langle J^P,\beta|\bm L^2|J^P,\alpha\rangle$ is also indicated. The two $1/2^\pm$ states are
\begin{align}
\label{heli1o2}
|1/2^-;1\rangle &= \sqrt{\frac{2}{3}}\,|^20_{1/2}\rangle - \sqrt{\frac{1}{3}}\,|^42_{1/2}\rangle \quad \textrm{with}\quad w_{11}=2,\\
|1/2^+;1\rangle &= \sqrt{\frac{2}{3}}\,|^21_{1/2}\rangle - \sqrt{\frac{1}{3}}\,|^41_{1/2}\rangle \quad \textrm{with}\quad w_{11}=2. 
\end{align}
It is possible to write $w_{11}=l_\textrm{eff}(l_\textrm{eff}+1)$ where $l_\textrm{eff}$ is a natural number and can be considered as an effective orbital angular momentum. $1/2^\pm$ states are then characterized by $l_\textrm{eff}=1$. 
There are two $3/2^-$ states 
\begin{align}
\label{heli3o2P}
|3/2^-; 1\rangle &= \sqrt{\frac{2}{3}}\,|^22_{3/2}\rangle + \sqrt{\frac{1}{6}}\,|^40_{3/2}\rangle - \sqrt{\frac{1}{6}}\,|^42_{3/2}\rangle, \\
|3/2^-; 2\rangle &= \sqrt{\frac{1}{2}}\,|^40_{3/2}\rangle + \sqrt{\frac{1}{2}}\,|^42_{3/2}\rangle, 
\end{align}
which are coupled with the matrix
\begin{equation}
\label{mat}
w_{\alpha\beta} =
\begin{pmatrix}
5 & -\sqrt{3} \\
-\sqrt{3} & 3
\end{pmatrix}.
\end{equation}
For this matrix, it is not possible to assign a value of $l_\textrm{eff}$ neither to $w_{11}$ nor to $w_{22}$. But its eigenvalues are $2$ and $6$, which correspond to $l_\textrm{eff}=1$ and $2$ respectively. There are also two $3/2^+$ states 
\begin{align}
\label{heli3o2m}
|3/2^+;1\rangle &= \sqrt{\frac{2}{3}}\,|^21_{3/2}\rangle + \sqrt{\frac{1}{30}}\,|^41_{3/2}\rangle + \sqrt{\frac{3}{10}}\,|^43_{3/2}\rangle, \\
|3/2^+;2\rangle &= \sqrt{\frac{9}{10}}\,|^41_{3/2}\rangle + \sqrt{\frac{1}{10}}\,|^43_{3/2}\rangle, 
\end{align}
which are coupled with the same matrix~(\ref{mat}). So states $3/2^-$ and $3/2^+$ are degenerate in our model. 

\begin{table}[htb]
\begin{center}
    \begin{tabular}{cccll}
        \hline\hline
 $J^P$ & $n_r$ & $l_\text{eff}$ & $cccg$ & $bbbg$ \\
\hline
$1/2^\pm$ & 0 & 1 & 1.842 & 1.784 \\
$3/2^\pm$ & 0 & 1 &  1.842  &1.784 \\
$3/2^\pm$ & 0 & 2 & 2.350  & 2.336 \\
$1/2^\pm$ & 1 & 1 & 2.552  & 2.469 \\
$3/2^\pm$ & 1 & 1 & 2.552 & 2.469 \\
$3/2^\pm$ & 1 & 2 & 2.938  & 2.880 \\
        \hline\hline
    \end{tabular} 
    \caption{Mass gap $m_{H\!B}-m_B$ in GeV for the lowest $J=1/2$ and $J=3/2$ $cccg$ and $bbbg$ hybrid baryons for a gluon with a helicity. The number $n_r = 0 \,(1)$ indicates the ground state (the first radial excitation) for the given values of $J^P$. \label{tab:mqqqgh}}
\end{center}
\end{table}

The comparison between Table~\ref{tab:mqqqgs} and Table~\ref{tab:mqqqgh} shows that the ground states contain  $1/2^-$ states in both cases but that the mass gaps differ. Moreover, the hierarchy of the states is also very different. In particular, positive and negative parity states are degenerate in Table~\ref{tab:mqqqgh}.  This demonstrates that the assignation of a spin or a helicity to the gluon has significant implications on the properties of hybrid baryons. As more physical results are obtained with the helicity status for two-gluon glueball, we also consider that the relevant results are presented in Table~\ref{tab:mqqqgh} and not in Table~\ref{tab:mqqqgs}. The main results that can be taken from Table~\ref{tab:mqqqgh} are:
\begin{itemize}
  \item The hierarchy is the same for $cccg$ and $bbbg$ states.
  \item The common value $l_\textrm{eff} =1$ for $1/2^\pm$ and some $3/2^\pm$ states causes their degeneracy. We checked that the same results are obtained by working directly with eigenstates of the operator $\bm L^2$.
  \item The lowest states have $J^P=1/2^\pm$ and $3/2^\pm$, and have a common mass around 1.8 GeV above the one of the ground state baryon.
  \item As in the case of $gg$ systems, no states with $l_\textrm{eff} = 0$ exist \cite{math08}.
\end{itemize}
For reasons that we present in Sec.~\ref{sec:conclu}, it is not obvious to compare these masses computed in the heavy sector with masses obtained in the light sector, which is the target of future experiments. For instance, in the lattice QCD work \cite{dude12} with computations performed with $m_\pi=396$~MeV, only positive parity spectra are presented. $1/2^+$ and $3/2^+$ hybrid baryons have masses close to each others and the hybrid-$\Delta$ is around 1.5 GeV above the baryon $\Delta$. Some similarities exist but an extension of our model to the light sector is needed to draw reliable conclusions. Let us remind that the interactions considered in this work are purely central. For instance, light $J_B^{P_B}=1/2^+$ and $3/2^+$ baryons are degenerate with our Hamiltonian~(\ref{HB}). With spin dependent contributions some degeneracies would be raised. 

The heavy hybrid baryons computed have quantum numbers compatible with ordinary baryons. So, the question of mixing between these configurations can be examined. In a large-$N_c$ approach of hybrid baryons, it appears that this mixing vanishes in the heavy quark limit \cite{chow99}. Assuming this result is valid in the real world with $N_c = 3$, it seems then reasonable that such mixing has very small effects for $cccg$ and $bbbg$ states and can be ignored.

\section{Concluding remarks and outlook}
\label{sec:conclu}

In this work, a heavy hybrid baryon is considered as a two-body system composed of a quark core and a gluon interacting via a simple QCD-inspired potential, a picture similar to the quark-diquark approximation for baryons. The dominant interaction is a potential simulating the flux tube confinement, and the Casimir scaling is assumed to differentiate interactions between triplet or octet color sources. A one-gluon exchange potential is also added. The main ingredient of this model is that the helicity of the gluon is correctly taken into account. Masses of $cccg$ and $bbbg$ hybrid baryons are computed by assuming the quark core in its ground state. Results obtained seem reasonable and are by some aspects similar to the ones obtained for light hybrid baryons computed within a lattice QCD calculation. So, we consider this paper as a proof of concept which can be extended for the study of light hybrid baryons, more interesting from an experimental point of view. But three aspects of our model must be improved.

First, it is necessary to use a universal potential model that can provide good spectra for ordinary and exotic hadrons. The seminal works \cite{godf85,caps86} can be a good starting point. The semirelativistic Hamiltonians developed in these papers contain relativized potentials and sophisticated spin contributions. An improved version with a screening effect for the linear confinement has been recently proposed \cite{weng24}. Using the Casimir scaling, a version for glueballs and hybrid hadrons could be tested. 

Second, the formation of a diquark inside a baryon is favored by a strong mass asymmetry between the quarks or the presence of a high angular momentum \cite{flec88}. This indicates that the formation of a compact three-quark cluster is probably not favored in ground states of light hybrid baryons. We nevertheless think that this difficulty can be overcome by allowing the quark core to be in a superposition of different states with a mixing controlled by the dynamics of the gluon. The coupling interaction could be computed as a perturbation arising from the difference between the full four-body Hamiltonian and the quark-core gluon one. 

Third, if the mixing between hybrid and ordinary baryons is suppressed in the heavy quark limit \cite{chow99}, the question must be seriously considered for light quarks. However, in a lattice QCD work \cite{dude12}, it is remarked that hybrid states have been easily identified within a dense spectrum of $qqq$ states. This is a possible indication that the mixing is not as strong as expected. Nevertheless, within a potential model, a scheme to treat this problem can be used which could be similar to the one developed to treat the quark-diquark mixing in the quark core.

An analytic scheme to study the phenomenology of hadrons, whose connection with QCD is clearly stated, can be obtained starting from the large number of colors ($N_c$) limit of QCD \cite{thoo74,witt79}. It is particularly fruitful in the light baryon sector. This approach can be combined with potential models to gain new insights into the structure of hadrons \cite{buis10,buis12,will16,buis22}. So, we plan to use a combined potential and Large-$N_c$ approach to study light hybrid baryons. It will then be possible to track down the properties of hybrid baryons from large values of $N_c$ to $3$, the value for the physical world. But a difficulty appears because the color wave function of the quark core has a mixed symmetry of the form $[2\,1\,\ldots\,1]$. This means it is, for instance, symmetrical under $1\leftrightarrow 2$ and antisymmetrical under $1\leftrightarrow 3,4,\ldots, N_c$. It is a priori not possible to build a totally antisymmetrical wave function above $N_c=3$, since there are not enough spin-flavor quark states. Except, if we work in the so-called Venezanio limit with a great number of flavors ($N_f$) \cite{vene79}, where $N_c\to \infty$, $N_f\to \infty$, and the ratio $N_c/N_f$ stays finite. Another difficulty is that the quark core becomes then a $N_c$-body system. This quantum many body problem can be solved with the envelope theory \cite{chev22,cimi24} which relies on the known exact solutions for the many-body harmonic oscillator Hamiltonian \cite{cint21}. Within this method, the computational cost is independent from the number of particles and the accuracy reached is sufficient to obtain relevant results for the baryon spectroscopy \cite{buis10,buis12,will16,buis22}.
 
\begin{acknowledgments}
L.C. would thank the Fonds de la Recherche Scientifique - FNRS for the financial support. This work was also supported under Grant Number 4.45.10.08. All authors would thank Cyrille Chevalier for providing accurate masses for the quark-cores. All authors would also thank the anonymous referee for his careful reading of our text and many relevant remarks, as well as the detection of a serious misprint.
\end{acknowledgments} 

\appendix

\section{Spin-helicity two-body states}
\label{sec:SpinHeli}

\subsection{Helicity states}

A particle with mass $m$, spin $s$, and helicity $\lambda$, moving in the direction specified by the polar angles $(\theta,\phi)$ with momentum magnitude $p$, is described by a helicity state $\ket{p\,\theta\,\phi;s\,\lambda}$ or $\ket{\bm{p};s\,\lambda}$. Helicity is defined as the projection of the spin along the direction of momentum and can take the following values depending on the mass of the particle

\begin{equation}
  \lambda =
    \begin{cases}
      -s,-s+1,\dots,s-1,s & \text{if $m \neq 0$}\\
      \pm s & \text{if $m=0$}
    \end{cases}.       
\end{equation}
The set of helicity states with all allowed values of $\lambda$ and $\bm{p}$ forms a complete set, known as the helicity basis \cite{jaco59}. A helicity state can be constructed from a state with a reference $4$-momentum $\Bar{p}$ by applying a Lorentz boost along the $z$-axis, denoted $L_z(\chi)$ with $\chi$ being the rapidity, followed by a suitable rotation $R(\alpha,\beta,\gamma) = e^{-i\alpha J_z}e^{-i\beta J_y}e^{-i\gamma J_z}$, where $(\alpha,\beta,\gamma)$ are the Euler angles and $\bm{J} = (J_x,J_y,J_z)$ are the angular momentum operators \cite{martin}

\begin{equation}\label{hs1}
    \ket{p\,\theta\,\phi;s\,\lambda} = R(\phi,\theta,-\phi)L_z(\chi)\ket{\Bar{p};s\,\lambda}.
\end{equation}
The reference $4$-momentum is typically chosen as $\Bar{p}=(m,0,0,0)$ for massive particles, and $\bar{p} = (1,0,0,1)$ for massless particles.

A helicity state with opposite momentum $-\bm{p}$, corresponding to polar angles $(\pi-\theta,\pi+\phi)$, can be constructed as follows
\begin{equation}
    \ket{-\bm{p};s\,\lambda} = (-1)^{s-\lambda}R(\phi,\theta,-\phi)e^{-i \pi J_y}L_z(\chi)\ket{\bar{p};s\,\lambda},
\end{equation}
where the rotation along the $y$-axis ensures the momentum has opposite angles to the original state, and the phase $ (-1)^{s-\lambda}$ is added for convenience \cite{jaco59}. Using this relation, a two-body helicity state in the center-of-mass frame can be constructed

\begin{equation}\label{hs2}
    \ket{\bm{p};s_1\lambda_1 s_2\lambda_2} = \ket{\bm{p};s_1\lambda_1} \otimes \ket{-\bm{p};s_2\lambda_2},
\end{equation}
since only particles carry momentum, not the interaction, in a potential model. The state \eqref{hs2} has a well-defined relative momentum $\bm{p}$ but not a total angular momentum, as it is not an eigenstate of $\bm{J}^2$ and $J_z$. The following state

\begin{equation}\label{jm}
    \ket{p;JM;\lambda_1 \lambda_2} = \left(\frac{2J+1}{4\pi}\right)^{1/2} \int \mathrm{d}\Omega \,D^{J*}_{M\,\lambda_1 - \lambda_2} (\phi,\theta,-\phi)\ket{p\,\theta\,\phi;s_1\lambda_1s_2\lambda_2},
\end{equation}
with $\mathrm{d}\Omega = \sin{\theta} \mathrm{d}\theta \mathrm{d}\phi$ and $D^J_{MM'}(\alpha,\beta,\gamma)$ the Wigner $D$-matrices, transforms as a state with total angular momentum $J$ and projection $M$

\begin{equation}
    R(\alpha,\beta,\gamma)\ket{p;JM;\lambda_1 \lambda_2} = \sum_{M'} D^J_{M'M}(\alpha,\beta,\gamma)\ket{p;JM';\lambda_1 \lambda_2}.
\end{equation}
The coefficient in front of the integral is chosen so that the state has a standard Lorentz-invariant normalization 

\begin{equation}
    \bra{p';J'M';\lambda'_1 \lambda'_2}\ket{p;JM;\lambda_1 \lambda_2} = \frac{4W}{p}\delta(p'-p)\delta_{JJ'}\delta_{MM'}\delta_{\lambda_1\lambda_1'}\delta_{\lambda_2\lambda_2'},
\end{equation}
with $W = (p^2+m_1^2)^{1/2} + (p^2+m_2^2)^{1/2}$ being the total energy. Note that the state $\ket{p;JM;\lambda_1 \lambda_2}$ still has well-defined individual masses $m_i$ and spins $s_i$, but these are omitted in the notation for brevity. From the properties of the Wigner $D$-matrices, the following selection rule emerges

\begin{equation}\label{sr}
    J \geq |\lambda_1 - \lambda_2|.
\end{equation}

\subsection{Canonical states}

One- and two-particle states can also be described in the canonical basis $\{\ket{\bm{p};s\,\mu}_c\}$, which is defined as an eigenstate of the operator $J_z$ with eigenvalue $\mu$. Starting from the reference state $\ket{\Bar{p};s\,\mu}$, a canonical state is constructed as follows

\begin{equation}\label{cs1}
    \ket{p\,\theta\,\phi;s\,\mu}_c = R(\phi,\theta,-\phi)L_z(\chi)R^{-1}(\phi,\theta,-\phi)\ket{\Bar{p};s\,\mu}.
\end{equation}
A two-body canonical state $\ket{p\,\theta\,\phi;s_1 \mu_1 s_2 \mu_2}_c$ is then built in a similar fashion to \eqref{hs2}. A total angular momentum $J$ is provided to the canonical state by using the usual $L-S$ coupling

\begin{equation}\label{jls}
    \ket{{}^{2S+1} L_J} = \sum_{\mu_1,\mu_2} (L\,m_L\,S\,m_S|J\,M) (s_1 \,\mu_1\, s_2 \,\mu_2|S\,m_S) \int \mathrm{d}\Omega \, Y^L_{m_L}(\Omega)\ket{p\,\theta\,\phi;s_1 \mu_1 s_2 \mu_2}_c,
\end{equation}
where $(a\,b\,c\,d|e\,f)$ is a Clebsh-Gordan coefficient and $Y^l_m(\Omega)$ is a spherical harmonics. The transformation from the helicity basis to the canonical basis is given by

\begin{equation}\label{trans}
    \ket{JM;\lambda_1 \lambda_2} = \sum_{S,L} \left(\frac{2L+1}{2J+1}\right)^{1/2} (L\,0\,S\,\lambda_1-\lambda_2|J\,\lambda_1-\lambda_2)(s_1\,\lambda_1\,s_2 \,- \lambda_2|S\,\lambda_1-\lambda_2) \ket{{}^{2S+1} L_J},
\end{equation}
where the sum runs over all values of $L$ and $S$ such that $|s_1-s_2|\leq S \leq s_1 + s_2$ and $|J-S| \leq L \leq J+S$.

\subsection{Parity}

Helicity states, as defined in equation \eqref{jm}, are not eigenstates of the parity operator, as it should be for a physical state of the strong interaction. This requirement is fulfilled by the following linear combination, with the same notation as in \cite{math08}

\begin{equation}\label{parity}
    \ket{H_\pm;J^P;\lambda_1\lambda_2} = \frac{1}{\sqrt{2}} \left[\ket{JM;\lambda_1 \lambda_2} \pm \ket{JM;-\lambda_1-\lambda_2}\right],
\end{equation}
which has the parity eigenvalue given by
\begin{equation}\label{P}
    P = \pm \eta_1\eta_2 (-1)^{J-s_1-s_2},
\end{equation}
where $\eta_i$ denotes the intrinsic parity of the particle $i$. Each state are now characterized by the quantum numbers $J^P$.

\subsection{Helicity states for hybrid baryons}

In our quark core model, a hybrid baryon can be modeled as a two-body system consisting of a massive quark core with spin $J_C$ and parity $P_C$, and a massless gluon with helicity $\lambda_g = \pm 1$ and parity $P_g=-1$. Following the discussion of Sec. \ref{sec:OC}, the two body helicity states are built for $J_C = 1/2$ and a positive parity quark core. Considering the selection rule on the total angular momentum \eqref{sr} and the parity eigenvalue \eqref{P}, the hybrid baryon is described by a basis of four states

\begin{subequations}
\begin{align}
    & \ket{H_+;\left(k+\frac{1}{2}\right)^P;\frac{1}{2}1} \text{ with } P =  (-1)^k \Rightarrow \frac{1}{2}^+,\frac{3}{2}^-,\frac{5}{2}^+,\dots \\
    & \ket{H_-;\left(k+\frac{1}{2}\right)^P;\frac{1}{2}1} \text{ with } P =  -(-1)^k \Rightarrow \frac{1}{2}^-,\frac{3}{2}^+,\frac{5}{2}^-,\dots \\
    & \ket{H_+;\left(k+\frac{3}{2}\right)^P;-\frac{1}{2}1} \text{ with } P =  -(-1)^k \Rightarrow \frac{3}{2}^-,\frac{5}{2}^+,\dots \\
    & \ket{H_-;\left(k+\frac{3}{2}\right)^P;-\frac{1}{2}1} \text{ with } P =  (-1)^k \Rightarrow \frac{3}{2}^+,\frac{5}{2}^-,\dots
\end{align}
\end{subequations}
Here, the allowed quantum numbers $J^P$ for each state are indicated on the right. The decomposition into canonical states \eqref{jls} is given below, following the basis transformation \eqref{trans}

\begin{subequations}\label{hel_HB}
\begin{align}
    \ket{H_+;J^P;\frac{1}{2}1} & = \sqrt{\frac{2}{3}}\ket{{}^2 {k+1}_{J}}+\sqrt{\frac{k}{2(2k+1)}}\ket{{}^4 {k-1}_{J}} -\sqrt{\frac{k+2}{6(2k+1)}}\ket{{}^4 {k+1}_{J}}, \label{hel_HB_1}\\
    \ket{H_-;J^P;\frac{1}{2}1} & = \sqrt{\frac{2}{3}}\ket{{}^2 {k}_{J}}+\sqrt{\frac{k}{6(2k+3)}}\ket{{}^4 {k}_{J}} -\sqrt{\frac{k+2}{2(2k+3)}}\ket{{}^4 {k+2}_{J}}, \label{hel_HB_2}\\
    \ket{H_+;J^P;-\frac{1}{2}1} & = \sqrt{\frac{k+3}{2(2k+3)}}\ket{{}^4 {k}_{J}} +\sqrt{\frac{3(k+1)}{2(2k+3)}}\ket{{}^4 {k+2}_{J}}, \label{hel_HB_3}\\
    \ket{H_-;J^P;-\frac{1}{2}1} & = \sqrt{\frac{3(k+3)}{2(2k+5)}}\ket{{}^4 {k+1}_{J}} +\sqrt{\frac{k+1}{2(2k+5)}}\ket{{}^4 {k+3}_{J}}. \label{hel_HB_4}
\end{align}
\end{subequations}
It is straightforward to verify that all these states are orthonormal. Since canonical states are eigenstates of the operators $\bm{J}^2,\bm{L}^2$ and $\bm{S}^2$, the mean values of these operators can be computed for each of the above helicity states. Results are presented in Table \ref{tab:helicity_operators}.

Similar calculations for the helicity states and mean values have been performed for $J_C=3/2$.

\begin{table}[htb]
\centering
\begin{tabular}{ccccc}
    \hline
    \hline
    $\hat{Q}$ & $\bm{J}^{2}$ & $\bm{L}^{2}$ & $\bm{S}^{2}$ & $\bm{L}\cdot\bm{S}$ \\
    \hline
    $\ket{H_+;J^P;\frac{1}{2}1}$ & $J(J+1)$ & $J(J+1)+5/4$ & $7/4$ & $-3/2$ \\
    $\ket{H_-;J^P;\frac{1}{2}1}$ & $J(J+1)$ & $J(J+1)+5/4$ & $7/4$ & $-3/2$ \\
    $\ket{H_+;J^P;-\frac{1}{2}1}$ & $J(J+1)$ & $J(J+1)-3/4$ & $15/4$ & $-3/2$ \\
    $\ket{H_-;J^P;-\frac{1}{2}1}$ & $J(J+1)$ & $J(J+1)-3/4$ & $15/4$ & $-3/2$ \\
    \hline
    \hline
\end{tabular}
    \caption{Mean values of different operators $\hat{Q}$ for the helicity states \eqref{hel_HB}.}
    \label{tab:helicity_operators}
\end{table}

\section{Lagrange-mesh method for two-body systems}
\label{sec:LagMesh}

The two-body system is solved with the Lagrange-mesh method \cite{baye15} which is very easy to use and very accurate. For two particles with spin, the method is described in \cite{sema01}. For the quark core-gluon system, a general wave function is approximated by the following expansion
\begin{equation}
\label{psiLM}
| \psi\rangle = \sum_{\alpha=1}^{N_h} \sum_{i=1}^{N_{L\!M}} C_{i \alpha}\,| f_i; J^P,\alpha \rangle
\end{equation}
in the basis $\{ | f_i; J^P,\alpha \rangle\} = \{ | f_i\rangle | J^P,\alpha \rangle\}$, with $\langle J^P,\beta | J^P,\alpha \rangle =\delta_{\beta\alpha}$. A sum runs on the $N_h$ helicity channels $\{ | J^P,\alpha \rangle \}$ defining the state and another one runs on $N_{L\!M}$ Lagrange radial functions $\{ f_i\}$ such that
\begin{equation}
\label{fi}
\langle \bm r | f_i,J^P,\alpha \rangle = \frac{1}{\sqrt{h} r} f_i\left( \frac{r}{h} \right) | J^P,\alpha \rangle.
\end{equation}
These functions are associated with $N_{L\!M}$ dimensionless mesh points $\{x_i\}$, are orthonormal (at the Gauss approximation), and vanish at all mesh points but one. Coefficients $C_{i \alpha}$ are computed by diagonalizing the Hamiltonian matrix with elements $\langle f_j; J^P,\beta |H| f_i; J^P,\alpha \rangle$, and $h$ is the only non linear parameter fixing the scale of the system (the method is not very sensitive to the value of this parameter). 

When only one channel is present, the computation of the matrix elements is described in \cite{sema01}. The only difference is the replacement of the mean values $l(l+1) = \langle \bm L^2 \rangle$ by their helicity counterparts $\langle J^P,\alpha |\bm L^2| J^P,\alpha \rangle = w_{\alpha\alpha}$ (see Sec.~\ref{sec:MHHBheli}). Let us note that the computation of $\langle \sqrt{\bm p^2+m^2} \rangle$ first involves calculating the eigenvalues of the operator $\bm p^2+m^2$ in the basis, which is easy to perform. 

For some $J^P$ quantum numbers, two or more channels must be taken into account. In (\ref{psiLM}), the same number of mesh points and the same value of the scale parameter are chosen for all helicity channels. This is not compulsory but this considerably simplifies the computation of the non diagonal matrix elements thanks to the orthogonality condition on the functions $\{ f_i\}$. As the interaction is purely central, the coupling of helicity channels is only due to the operator $\bm L^2$ in $\bm p^2$. With $\beta \ne \alpha$, it is easy to show that
\begin{align}
\label{L2coupl}
&\langle f_j;J^P,\beta |\bm p^2 + m^2| f_i; J^P,\alpha \rangle =\frac{ w_{\beta\alpha}}{h^2 x_i^2} \delta_{ji} , \\
&\langle f_j;J^P,\beta |V(r)| f_i; J^P,\alpha \rangle = 0.
\end{align}

\end{document}